# An Educational Website on Interferometry

Veena Singh, Richa Dubey, P.K. Panigrahi[*], and K. Muralidhar

*Laser Technology Programme*

*Indian Institute of Technology Kanpur, Kanpur India*

[*]`panig@iitk.ac.in`

*Abstract -* The present work describes a website designed for remote teaching of optical measurements using lasers. It enables senior undergraduate and postgraduate students to learn theoretical aspects of the subject and also have a means to perform experiments for better understanding of the application at hand. At this stage of web development, optical methods considered are those based on refractive index changes in the material medium. The website is specially designed in order to provide remote access of expensive lasers, cameras, and other laboratory instruments by employing a commercially available web browser. The web suite integrates remote experiments, hands-on experiments and life-like optical images generated by using numerical simulation techniques based on Open Foam software package. The remote experiments are real time experiments running in the physical laboratory but can be accessed remotely from anywhere in the world and at any time. Numerical simulation of problems enhances learning, visualization of problems and interpretation of results. In the present work hand-on experimental results are discussed with respect to simulated results. A reasonable amount of resource material, specifically theoretical background of interferometry is available on the website along with computer programs image processing and analysis of results obtained in an experiment.

## I. INTRODUCTION

Optical methods of measurement are known to have specific advantages in terms of spanning a field-of-view and being inertia-free. In use for over half a century, optical methods have seen resurgence over the past decade. The main factors responsible are the twin developments in the availability of cost-effective lasers along with high performance computers. Laser measurements in thermal sciences have been facilitated additionally by the fact that fluid media are transparent and heat transfer applications in fluids are abundant. Whole-field laser measurements of flow and heat transfer in fluids can be carried out with a variety of configurations: shadowgraph, schlieren, interferometry, speckle and PIV, to name a few. Here, temperature field measurement in fluids by laser interferometry has been addressed.

The ability to record interferograms on a PC using CCD cameras has greatly simplified image analysis. It is possible to enhance image quality and perform operations such as edge detection and fringe thinning by manipulating numbers representing the image. Image analysis techniques have also been discussed in the present article.

When combined with holography, laser interferometry can be extended to map three dimensional fields. Holographic interferometry can be cumbersome in some applications due to the need of holographic plates, particularly when large regions have to be scanned. This difficulty is circumvented by using an analytical technique called tomography. Here the interferograms are viewed as projection data of the thermal field. The three dimensional field is then reconstructed by suitable algorithms. In principle, tomography can be applied to a set of projection data recorded by shadowgraph, schlieren, interferometry or any of the other configurations.

With the onset of the information age, the use of computers and high speed networks has become common. The internet has made it possible to transfer information at unprecedented rates and to bring education out of the classrooms and share it around the globe. Educational courses are easily available on the internet in the form of text and video lectures. However engineering education is possible only if it is supplemented by laboratory experience. This is possible with the help of internet. The first step in this direction was taken by Arpai et al. of University of Naples Italy as early as 1997. They developed a Remote Laboratory (RemLab) based on a website for remote teaching of experimental techniques [1]. The next step was the development of a remote-access control system that allows users to perform controlled experiments via internet. To illustrate the design a dc motor control module was used [2].

The development of virtual laboratories along with remote access leads to complete learning of the course. Virtual laboratories enable the students to perform experiments any time and from any place. This technology is cost effective compared to the cost and space required to set a real time laboratory. These laboratories enable the students to do experiments that are dangerous or long term when carried out in a physical environment. These provide a fully interactive ambient for the students to do an experiment any number of times and simultaneously analyse the result.

Our project - An educational website on interferometry is a combination of virtual laboratory with remote access. It enables the client (student) an access to a remote research laboratory offering physical experiments through web interface from any computer connected to the internet. In addition, it provides a means to visualize results through numerical simulation [3]. Experiments are accompanied by theory as resource material on the web page. It is specially designed for 4[th] year undergraduate and 1[st] year postgraduate students to let them be acquainted with the fundamentals of interferometry and its applications.

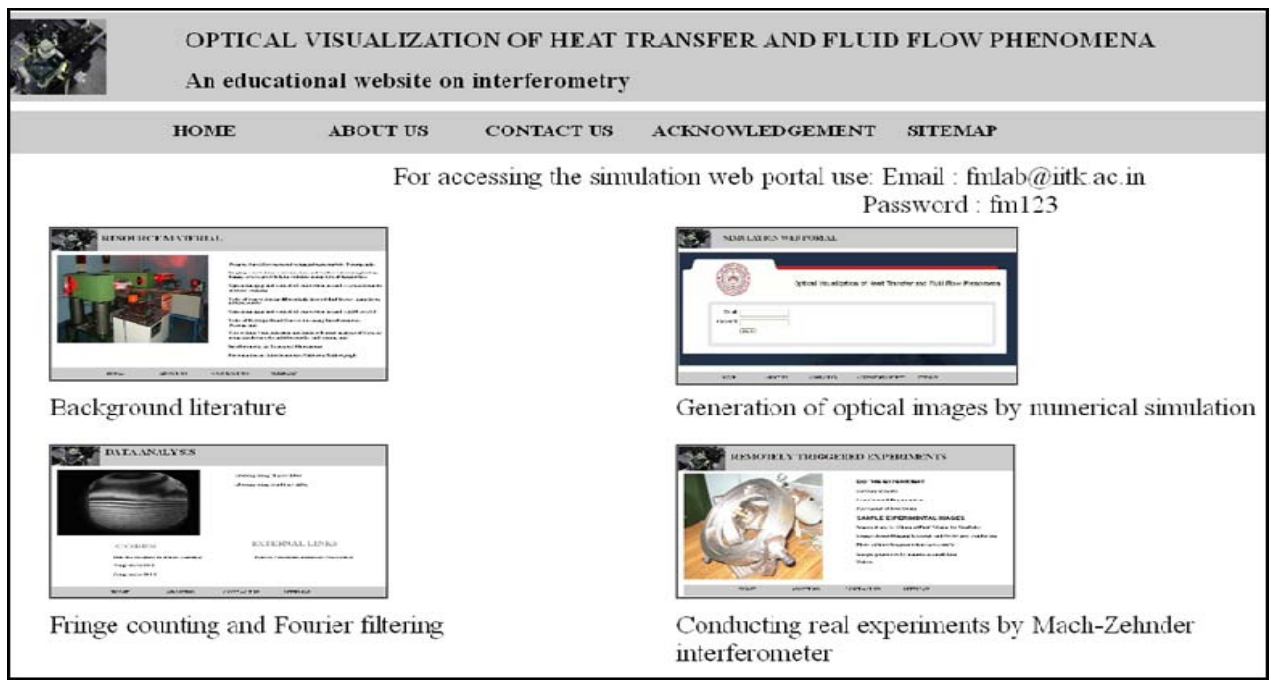

Fig. 1 Image of the opening page of the website

This website is designed to:

1. provide the learner the means to learn about interferometry, through the literature provided on site.

2. perform simulations and find the results via virtual lab and authenticate it with performing the real experiments via remote access;

3. search for novel ways to maintain student engagement at the same time as optimizing resources;

4. enable distance learning and accessing expensive laboratory experiments;

5. conduct training experiments.

## II. OVERALL DESCRIPTION OF THE WEBSITE

The website comprises four panes namely
  a. Background Literature and Introduction to the Subject
  b. Generation of Optical Images by Numerical Simulation
  c. Interferometric analysis, Fringe Counting and Fourier Filtering
  d. Conducting Real Experiments by Mach-Zehnder Interferometer

This website will be expanded in the future to include other categories of optical techniques.

## III. BACKGROUND LITERATURE

This part of the website provides the theoretical background on interferometry. It integrates material for beginners, research articles, doctoral thesis and presentations which provide better knowledge of the topic.

## IV. GENERATION OF OPTICAL IMAGES BY NUMERICAL SIMULATION

Optical techniques can be used for validating numerical solutions of the heat transfer equations and flow. In the present work, the three dimensional field of a dependent variable, say temperature is computed from a numerical solution of the applicable differential equations. Interferograms, schlieren, and shadowgraph are numerically generated based on the numerical solution and compared with those in the respective real experiments. Here, it is to be noted that interferograms yield information on temperature itself, schlieren for the first derivative, and shadowgraph for the Laplacian of the field variable. The data generated reveals to the students the simulation of optical patterns under various conditions of flow and heat transfer rates [6]-[8]. All the simulations are based on Open Foam software package and have been developed by *Tridiagonal Solutions,* Pune under the supervision of the authors. The governing equations are numerically solved for velocity (when flow is involved) and temperature. With temperature data available on a grid, optical patterns are determined.

The outcome is a one of the pane of the complete website, Optical Visualization of Heat transfer and Fluid Flow Phenomena, Figure 3. In the website, the source code and the executable version that solves flow and heat transfer problems are available. The code is general enough to treat the following situations:

  a. 1 and 2-dimensional geometries.
  b. Regular and irregular (multiply-connected) objects.
  c. Steady and unsteady, laminar incompressible flows with heat transfer
  d. Steady and unsteady classical heat conduction.
  e. Isothermal, constant heat flux and insulated surfaces.

f. Constant material properties – independent of temperature and pressure.
   g. Single phase transport phenomena

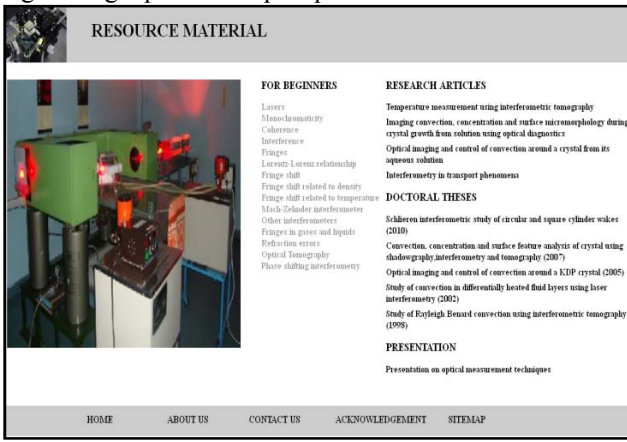

Fig. 2 Snapshot of background literature web portal

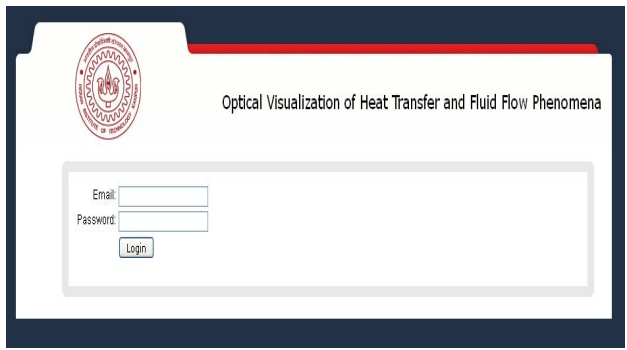

Fig. 3 Snapshot of simulation web portal

The user has a choice of coarse, medium, and fine meshes. Materials chosen can be copper and mild steel (conduction) and air as well as water (flow). The respective properties are available in a database. Once the solution for temperature is obtained, the equivalent optical images are internally calculated and displayed.

The flowchart for navigation through this web portal is shown in Figure 4.

Further, this portal of the web site is structured into three parts

### A. New Example

In this section user can solve examples of choice by navigating the web page. There are six examples which have been incorporated. These include

*1) Steady Heat Conduction:* In this example steady heat conduction in a solid body of rectangular and annular cross sections is simulated. The final results comprise temperature and temperature gradients shown as equivalent optical images (interferograms, schlieren and shadowgraph) and temperature contours. In general, the importance of boundary conditions is highlighted. Some results for rectangular geometry are shown in Figure 5. The left and right face of the rectangular geometry are kept at 350 K and 300 K respectively. The upper and lower faces are adiabatic. In this case shadowgraph is of no importance as it has zero value.

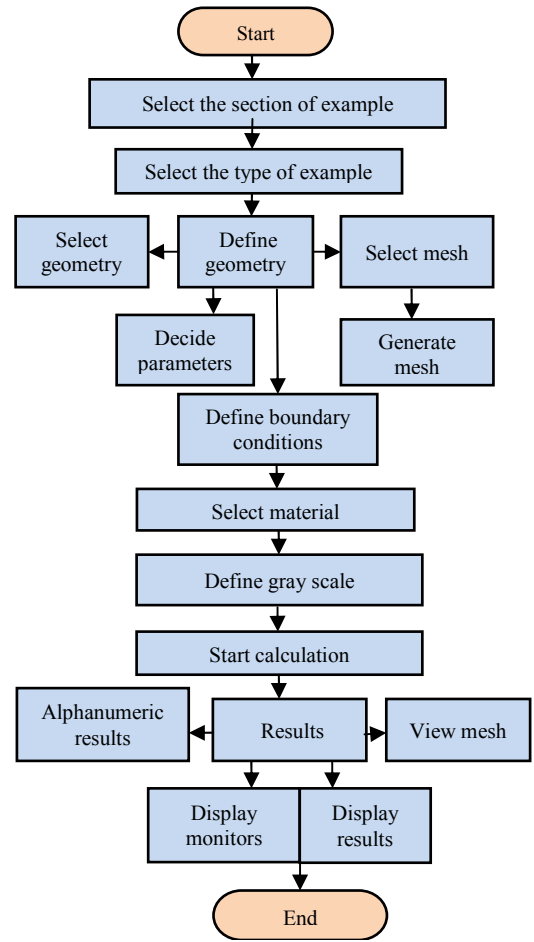

Fig. 4 Flowchart for navigation through the simulation web portal

*2) Unsteady Heat Conduction:* In this example unsteady heat conduction in a solid body of rectangular and annular cross sections is simulated. Here, the temperature field evolves from given initial conditions. The final results carry temperature and temperature gradients as their equivalent optical images (interferograms, schlieren and shadowgraph) and temperature contours. The role played by the boundary conditions is also of importance. Some results for annular geometry are shown in Figure 6. The outer and inner faces are kept at 400 K and 300 K respectively.

*3) Boundary Layer Flow over a Flat Plate:* In this example, flow over a flat plate with a sharp leading edge is simulated. The final results are in the form of temperature and temperature gradients as their equivalent optical images (interferograms, schlieren and shadowgraph), velocity and temperature contours. The major points of difference between laminar and turbulent profiles and the importance of Prandtl number are brought out in this example. Some results for this example are shown in Figure 7. The inlet temperature is 300 K and plate is kept at 320 K. Flow is laminar with a Reynolds number 1000 (per unit length) and Prandtl number 0.6.

4) *Buoyancy-driven Convection:* In this example buoyancy driven flow field, temperature distribution in an annulus and flat vertical plate are simulated for various thermal boundary conditions. The wall heat transfer rates are calculated from convection patterns. The numerically generated results for the velocity field and streamlines are presented. Parameters chosen should be such that the flow is laminar and steady (low Rayleigh numbers). Some results for buoyancy driven convection over a flat plate and in annular geometry are shown in Figure 8 and Figure 9. In case of convection over a vertical plate the ambient and the plate are at 300 K and 350 K respectively. In Figure 9 the outer and inner faces are kept at 300 K and 400 K.

5) *Vortex Shedding from a Cylinder:* Flow past a circular cylinder at low Reynolds numbers (when a clear vortex shedding regime is established) is studied. Flow is in the vertically upward direction. Velocity (instantaneous and time-averaged), and vorticity variation around the cylinder are available. The students can compare the separation process and vortex shedding phenomena with that available in the literature. The optical images obtained from a slightly heated fluid relative to the cylinder form the basis of comparison against experiments. Some results for this example are shown in Figure 10. Here, the inlet is at 300°K having a Reynolds number of 100 and the cylinder is kept at 350°K.

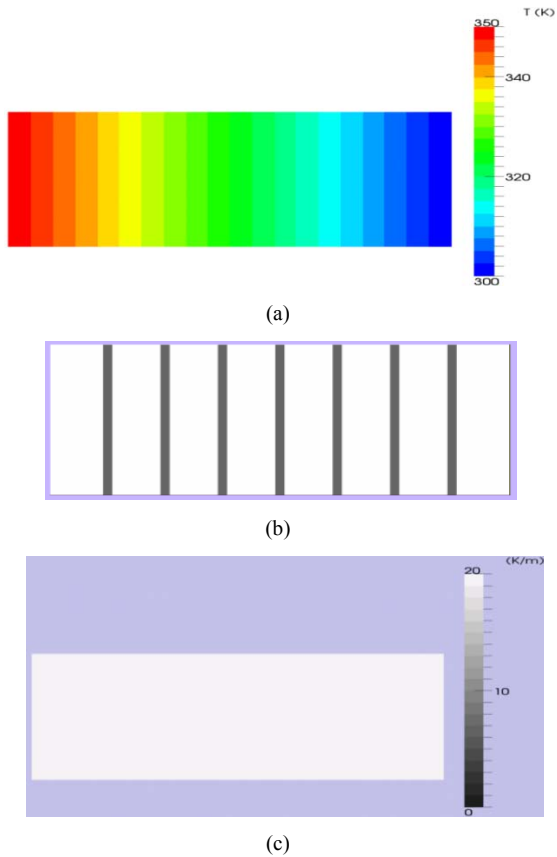

Fig. 5 Steady heat conduction in rectangular geometry (a) Temperature distribution (b) Interferometric fringes (c) Schlieren image of uniform intensity

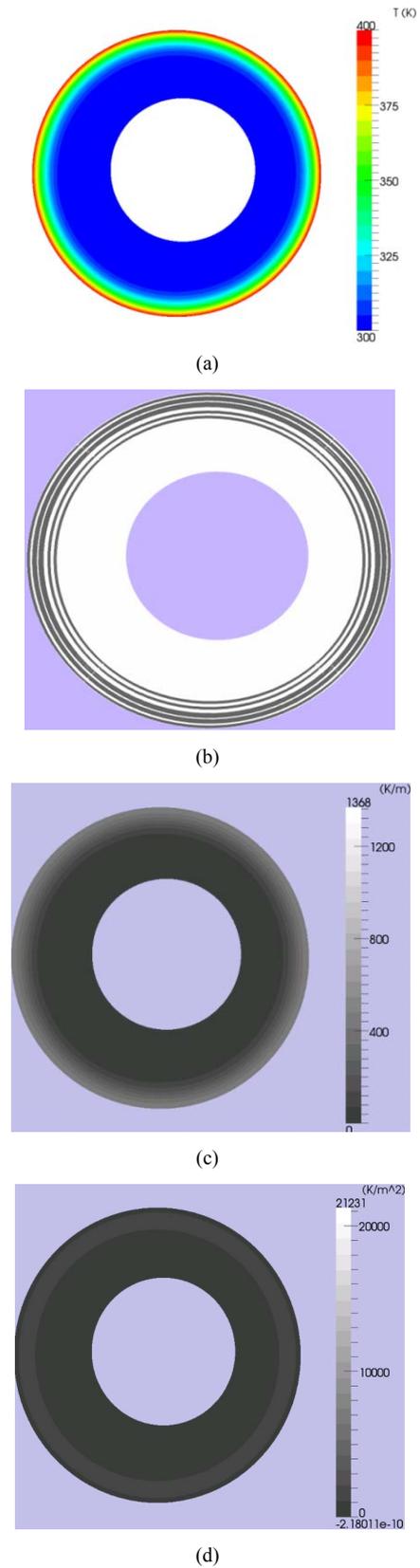

Fig. 6 Unsteady heat conduction for annular geometry (a) Temperature contours (b) Interferometric fringes (c) Schlieren image and (d) Shadowgraph

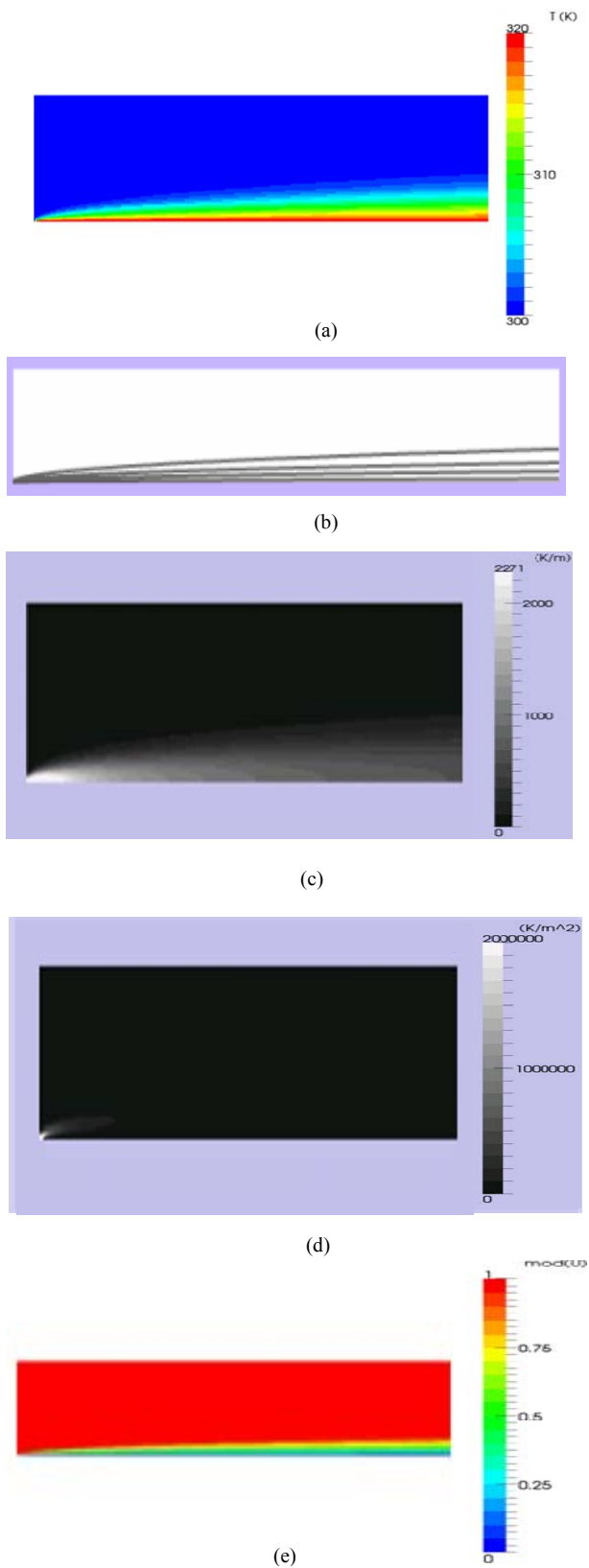

Fig. 7 Boundary layer flow (from left to right) over a flat plate (a) Temperature contours (b) Interferometric fringes (c) Schlieren intensity distribution (d) Shadowgraph (e) Velocity modulus.

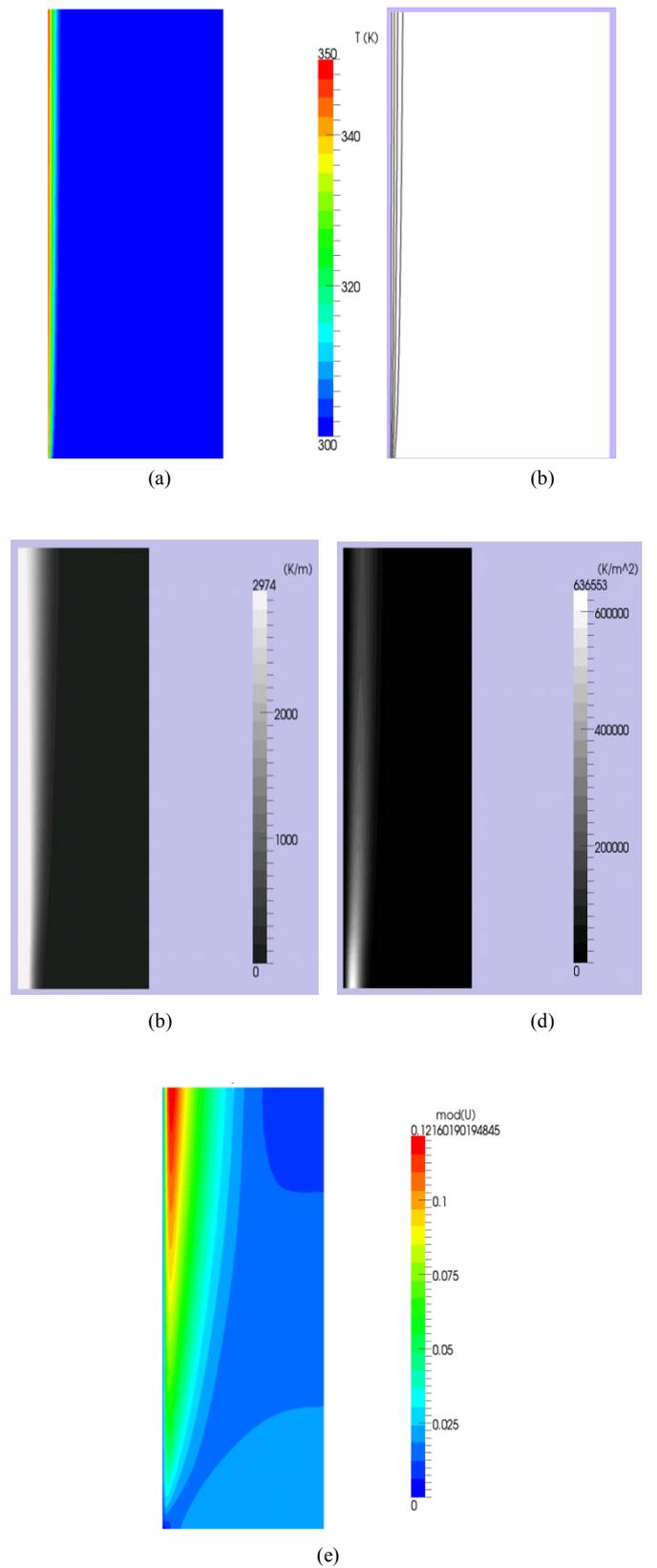

Fig. 8 Buoyancy driven convection over a vertical flat plate (a) Temperature contours (b) Interferometric fringes (c) Schlieren intensity distribution (zoomed image) (d) Shadowgraph (zoomed image) (e) Velocity modulus

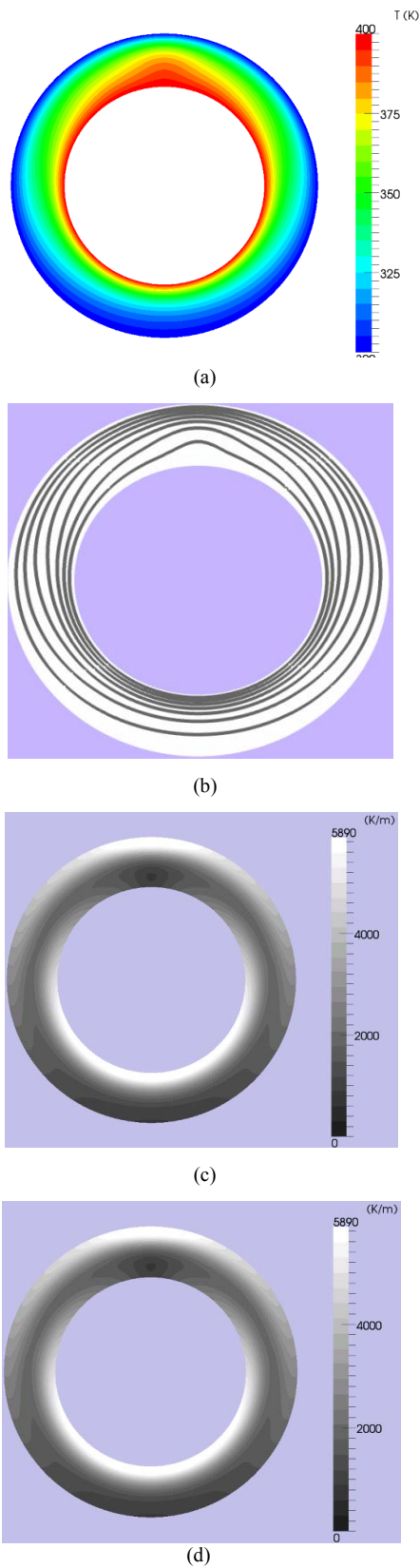

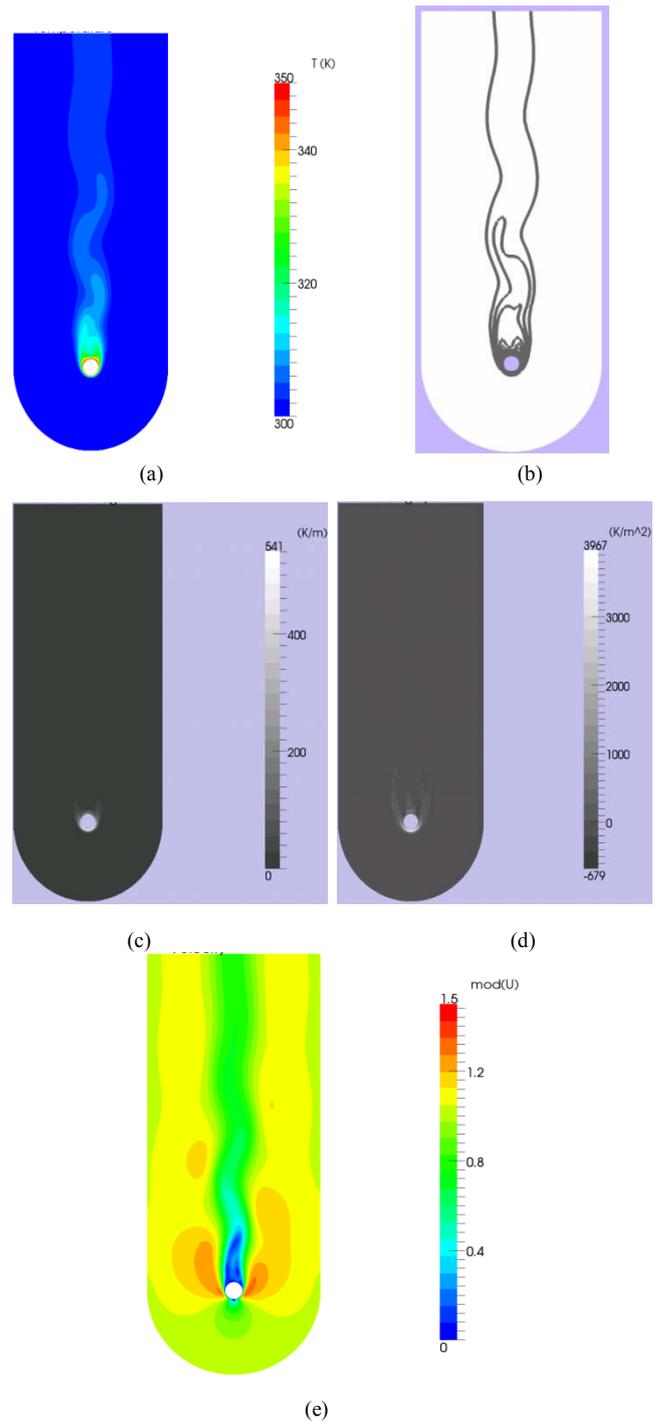

Fig. 9 Buoyancy driven convection in a annular geometry (a) Temperature contours (b) Interferometric fringes (c) Schlieren intensity distribution (d) Shadowgraph

Fig. 10 Vortex shedding from a circular cylinder (a) Temperature contours (b) Interferogram (c) Schlieren intensity distribution (d) Shadowgraph (e) Velocity modulus

6) *Flow and Heat Transfer from a Flat Surface with a Surface Mounted Rib:* This example is similar to boundary layer flow over a flat plate except that a rib of square cross-section is placed on it. The rib acts as a vortex generator and improves heat transfer from the surface. The initial condition can correspond to a hot plate placed in a cold fluid. The plate is then allowed to cool at a rate determined from the flow considerations. The cooling rate of the plate is monitored with and without the rib. Corresponding optical images reveal the wall heat flux as a function of time, approaching zero at steady state. Some results for this example are shown in Figure 11. Here, the inlet is at 300 K and the plate is kept at 350 K.

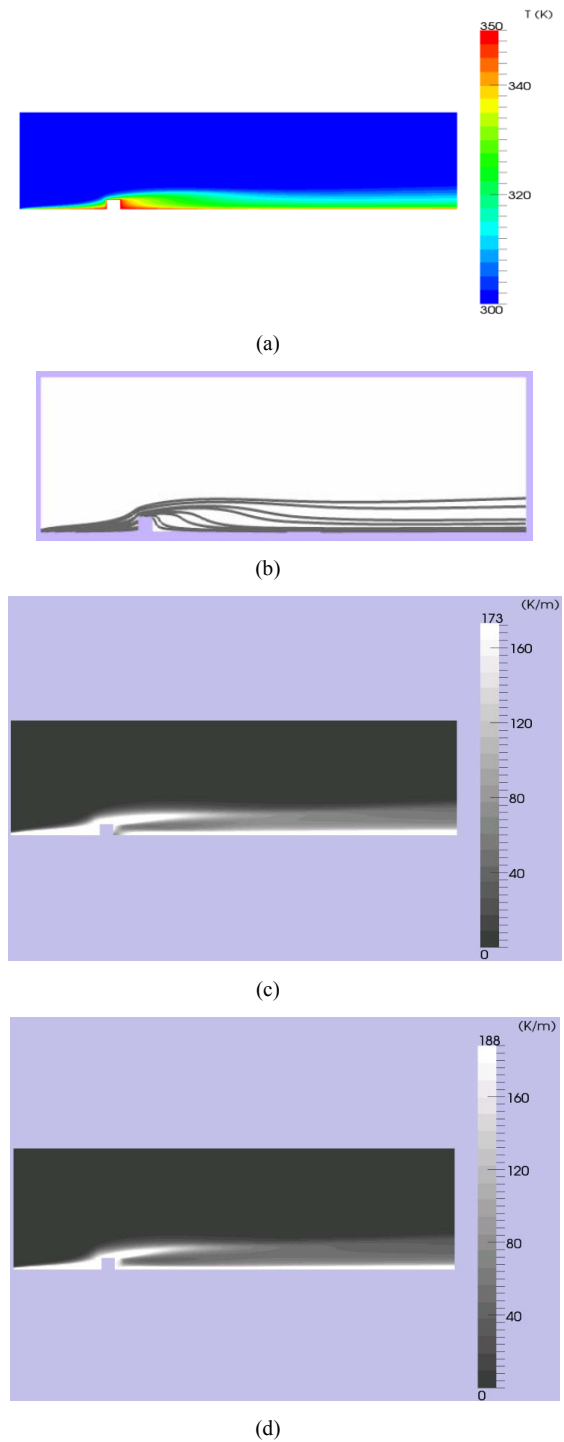

Fig. 11 Flow and heat transfer from a flat with a surface mounted rib (a) Temperature contours, (b) Interferograms (c) Schlieren intensity at Reynolds number of 10000(d) Schlieren intensity at Reynolds number of 25000

*B. My Example*

This section acts as a backup of the website. The examples simulated by the user are stored in this section so that user will be able to access those at any other point of time.

*C. Solved Example*

This is a demonstration section. It includes examples solved by the authors in advance for each of the six cases under different initial conditions to get best results.

V. FRINGE COUNTING AND FOURIER FILTERING

This pane mainly deals with the data analysis of the optical images. Optical images obtained in interferometry, using the CCD camera and A/D converters are noisy. Noise can be defined as undesired intensity variation which is superimposed over information. Noise is generated at different stages of the experiment: digitization, imperfection of the optical components, noise in the CCD array, and lastly the flow and thermal fields under study itself contribute to noise owing to edge effects, refraction and non uniform scattering and absorption. However, most authors consider, speckle to be a major source of noise in interferometry. The microscopic unevenness on the surface of the optical elements leads to diffraction. The superposition of diffraction patterns over the basic interference pattern that constitutes the signal is a speckle noise. The local fringe pattern generated by interference of diffracted rays corrupts the global intensity distribution. Thus speckle is characterized by high-wave number fluctuations (rapid transitions) in the intensity data. To retain the original intensity variation, images need to be processed so that noise can be removed and data can be extracted. Image processing operation in the interferometry is divided in to three categories: (a) Filtering (b) Image enhancement (c) and Thinning [4].

In the present work, filtering is of main concern. Filtering on some synthetic and experimental images has been illustrated in this pane. Basic programs needed to process the images obtained from the real experiments, some results obtained after processing and literature related to it are also included. The codes are provided in C as well as in Matlab. Students can download the programs any time and use them for the analysis of the interferograms. The snapshot of this web portal is shown in Figure 12.

VI. REMOTELY TRIGGERED EXPERIMENTS

A number of higher education institutions are establishing Remote Access Laboratories to empower the student as a self-learner and enhance the learning experience of those in engineering. Teachers obtain new possibility of completing theory from experiments without any investing extra time in a lesson. Students can take advantage of this web tool to become familiar with the possibilities of utilizing information technology in everyday life. The 4[th] portal of the website deals with performing a remotely triggered experiment. The snapshot of this portal is shown in Figure 13.

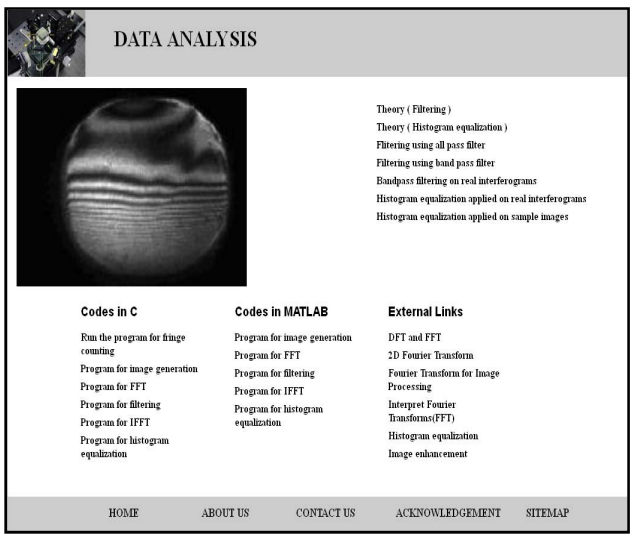

Fig. 12 Snapshot of the data analysis web portal

*A. Experimental Setup*

In the present work a setup is developed for the remote control of the Mach-Zehnder interferometer. It consists of an optical mount to hold the optics of the interferometer. It comprises the base plate, outer ring, inner ring, vertical movement hinge and horizontal movement hinge. A photograph of the optical mount is shown in Figure 14.

The base plate acts as a base of the optical mount. There are two base plates, inner and outer. The inner plate is of 140 mm length, 170 mm width and 15 mm thickness. The outer ring makes contact with inner ring exactly at a single point. This leads to two degrees of freedom of the optical mount, one about the horizontal axis and other about the vertical. The inner ring holds the optics. Both the rings are set on fine-thread screws so that there will be precise movement of optics without friction. The linear motion of the motor shaft is converted into rotary motion by means of fine-thread screws. The screws can by driven by a motor under computer control. The diameters of the outer and inner rings are about 151 mm and 191 mm respectively according to the corresponding mirror diameter. Hinges for vertical movement and horizontal movement are used to make contact between the inner and outer rings. There is a hinge connect at the periphery of both the rings to convert linear motion of the spokes into angular motion of the optic element such as a mirror or a beam splitter.

*B. Automation of the Optical Mount*

As the setup is used remotely, an electric DC stepper motor is employed to apply short pulses of torque. A stepper motor is a brushless, synchronous electromechanical device which converts electrical input pulses into discrete mechanical movements. When electrical command pulses are applied to it in the required sequence the shaft of a stepper motor rotates in the increment of discrete step. The rotation of motor is in proportion to the applied input pulses. The sequence of the applied pulses has a direct relationship with the direction of rotation of shaft. The length of rotation is governed by the number of input pulses applied and the frequency of the input pulses decides the speed of the motor shaft rotation.

Motor drivers are basically little current amplifiers. Their function is to take a low-current input control signal, and turn it into a proportionally higher-current output signal that can drive a stepper motor. Easy driver is used as a motor drive in the present work. It is permanently fixed to use 8 step micro stepping mode. It is based on the Allegro A3967 driver chip [9]. The image of the easy driver kit is shown in Figure 15.

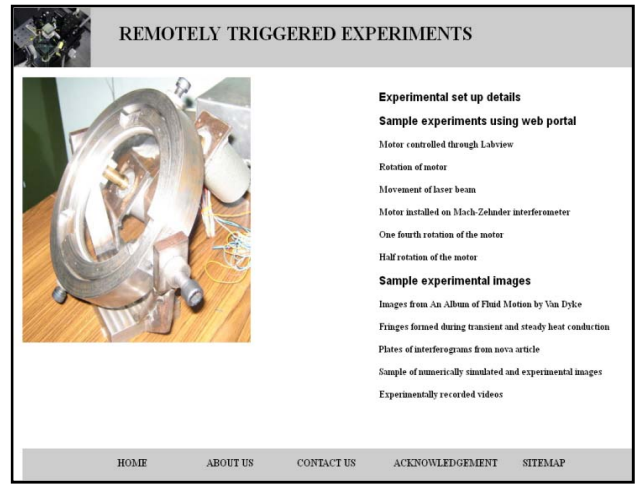

Fig. 13 Snapshot of the web portal for conducting remotely triggered experiments

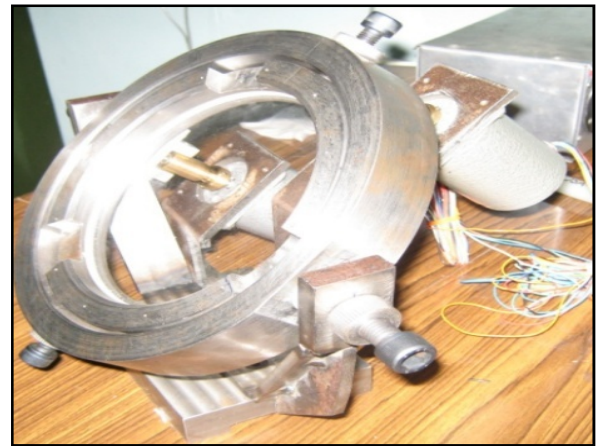

Fig. 14 Motorized optical mount for remote access of the interferometer

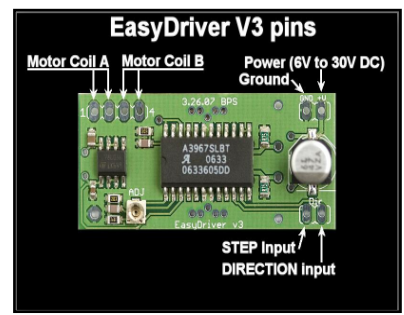

Fig. 15 Easy driver kit [9]

## C. Remote- Access Real Time Experiment

Remote-access laboratory is developed by the hardware framework described above with the execution and control supported in a LabVIEW environment. The front panel of the application can be published with the built-in web server in LabVIEW. A computer panel at the remote or host location is depicted in Fig. 16. Once the web server is enabled, LabVIEW generates front panel images of the VI and a URL that can be accessed from any web browser to link to the experiment [14]. Remotely triggered system architecture and the snapshots of the movement of laser beam via LabVIEW are shown in Figures 17 and 18 respectively.

## VII. CONCLUSIONS

Development of a web portal that is available remotely, 24 hours a day is reported. The present study shows the feasibility and the potential of such an educational website on the subject of laser interferometry for remote experiments on subject of heat transfer. Students have an access to a laboratory from anywhere through a personal computer connected to the internet. Remotely accessed experiments are a part of the webpage. Experiments are accompanied by theoretical material on optical methods on the web pages. The webpage is designed such that there could be further development with the ultimate aim of training students in optical measurement techniques in engineering. Theoretical support, tasks and numerical simulation are placed on the website link

http://202.3.77.50/~opticalv/interferometry/

next to experiments. We have also compared the optical images obtained from simulation against real experimental images. The four major aspects of the webpage are:

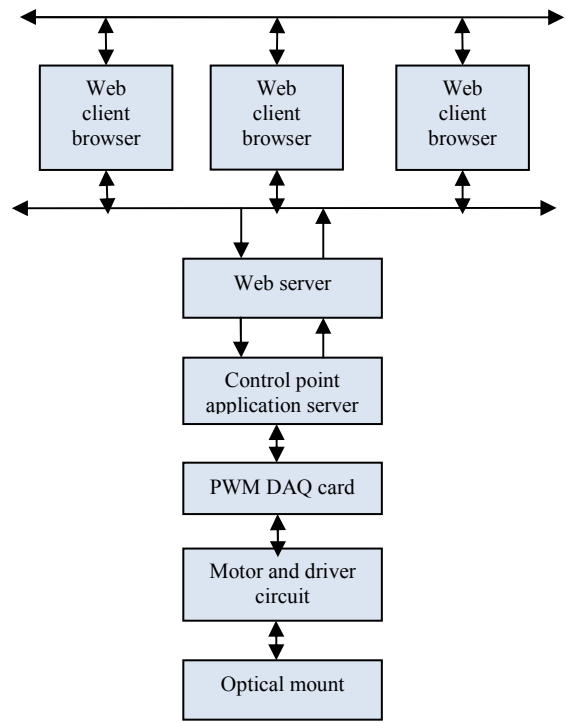

Fig. 17 Remotely triggered system architecture

a. Resource material on optical methods
b. Numerical simulation for generating optical images
c. Aspects of image processing
d. Remote trigger experiment

It is not claimed here that remote access laboratories are not a replacement of hands-on laboratories. They are intended for use in a limited way. They may, however, form the basis of collaborative research experiments across shared resources.

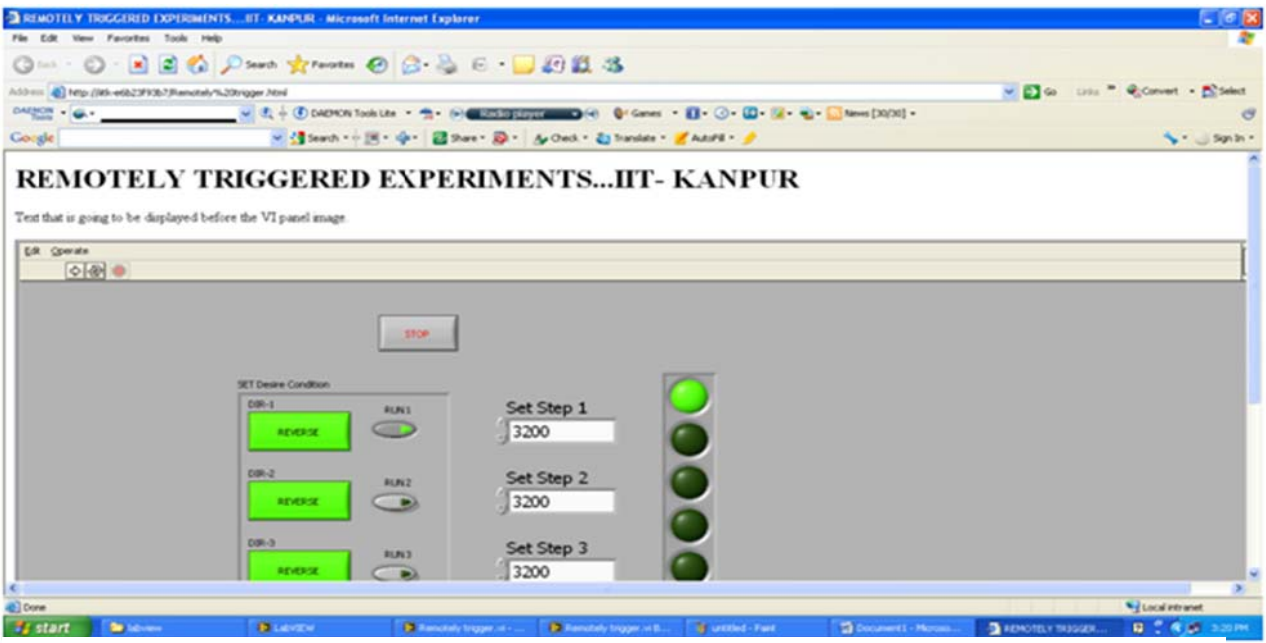

Fig. 16 Front panel of the VI at the remote server

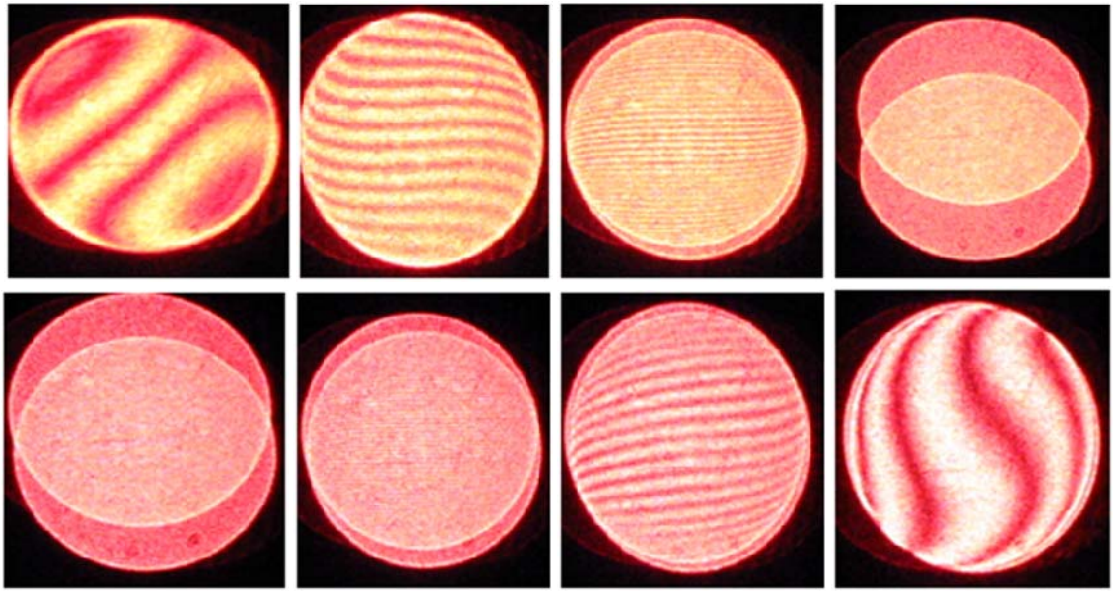

Fig. 18 Movement of laser beam controlled via LabVIEW over internet. The first row shows progressive misalignment while the second row shows improvement in alignment of a Mach-Zehnder interferometer

ACKNOWLEDGMENT

The authors thank the Ministry of Human Resource Development, New Delhi for financial support.